\documentclass[letterpaper,twocolumn,10pt]{article}
\pdfoutput=1

\usepackage{usenix-2020-09}

\usepackage{cite}

\usepackage[newfloat,frozencache=true,cachedir=.]{minted} 
\setminted{
    linenos=false,
    fontsize=\footnotesize,
    autogobble,
}

\usepackage{caption}
\newenvironment{code}{\captionsetup{type=listing}}{}
\SetupFloatingEnvironment{listing}{
name=Code,
placement=ht,
}

\usepackage{amsmath}
\usepackage[inline,shortlabels]{enumitem}
\usepackage[caption = false]{subfig}

\usepackage{todonotes}
\usepackage{menukeys}
\newcommand{\commentout}[1]{{}}

\newcommand{\frisbee}[1]{{\em Frisbee}}
\newcommand{\faultload}[1]{{\em Faultload}}
\newcommand{\manifest}[1]{{\em Manifest}}

\usepackage{booktabs}

\graphicspath{ {images/} }
\usepackage{epsfig}
\usepackage{hyperref}
\hypersetup{colorlinks,allcolors=black}

\begin{document}

\date{}

\title{\Large \bf Frisbee: automated testing of Cloud-native applications in Kubernetes}

\author{
{\rm Fotis Nikolaidis}\\
ICS-FORTH \\
fnikol@ics.forth.gr 
\and
{\rm Antony Chazapis}\\
ICS-FORTH \\
chazapis@ics.forth.gr 
\and
{\rm Manolis Marazakis}\\
ICS-FORTH \\
maraz@ics.forth.gr 
\and
{\rm Angelos Bilas}\\
ICS-FORTH \\
bilas@ics.forth.gr 
} 

\maketitle

\begin{abstract}
As more and more companies are migrating (or planning to migrate) from on-premise to Cloud, their focus is to find anomalies and deficits as early as possible in the development life cycle. We propose Frisbee, a declarative language and associated runtime components for testing cloud-native applications on top of Kubernetes. Given a template describing the system under test and a workflow describing the experiment, Frisbee automatically interfaces with Kubernetes to deploy the necessary software in containers, launch needed sidecars, execute the workflow steps, and perform automated checks for deviation from expected behavior. We evaluate Frisbee through a series of tests, to demonstrate its role in designing, and evaluating cloud-native applications; Frisbee helps in testing uncertainties at the level of application (e.g., dynamically changing request patterns), infrastructure (e.g., crashes, network partitions), and deployment (e.g., saturation points). Our findings have strong implications for the design, deployment, and evaluation of cloud applications. The most prominent is that: erroneous benchmark outputs can cause an apparent performance improvement, automated failover mechanisms may require interoperability with clients, and that a proper placement policy should also account for the clock frequency, not only the number of cores. 
\end{abstract}

\section{Introduction}
\label{sec:intro}
Cloud-native applications refer to an architectural style where applications consist of loosely coupled Microservices, that run on a containerized and dynamically orchestrated platform (with the help of technologies like Kubernetes, Docker, etc.), in dynamic environments provided by the public, private and hybrid clouds. This architecture results in significant application downtime reduction, dynamic scaling-up/scaling-down of cloud resource utilization as per business needs, and increased responsiveness to user demand. However, the benefits of cloud-native architectures do not come for free. Testing these applications requires a series of complex and time-demanding activities, from the deployment and configuration of the elastic system, to the simulation of normal and abnormal conditions, and the collection and analysis of distributed metrics. 

System testers have traditionally emphasized writing static test cases since most of them have not had the tooling or the infrastructure to run distributed scenarios. Unfortunately, these manual approaches are highly inadequate for testing cloud-native applications. Firstly, manual tests usually come late in the development cycle; when bugs are revealed, reworked, and the additional retesting cost is high and impacts the product release. Secondly, manual testing is biased. Testers know the server where the application resides, along with its CPUs, memory, and network bandwidth, and they can test against those expectations~\cite{abedi2017conducting,cavalli2015survey}. Thirdly, most test cases operate through distinct phases; setup, steady-state, and (optionally) shutdown. That requires synchronization between phases to ensure proper and consistent testing. However, the non-deterministic nature of distributed systems makes manual or time-based solutions prone to inconsistent behaviors, forcing engineers to spend their time chasing down issues that may not even exist. Finally, a golden rule that is hardly viable with manual methods is to create similar environments for dev, test, and production, so that the system under test should be as similar as possible, whether it runs on a developer's laptop, in a testing environment, or in production.

Kubernetes is a promising solution for unbiased grey-box testing of cloud-native applications. It provides a cheap and disposable environment where researchers can deploy the system under evaluation (SUT) and stress them against testing utilities, such as performance benchmarks~\cite{cooper2010benchmarking}, linearizability checkers~\cite{jepsen}, or Chaos Engineering tools~\cite{basiri2016chaos}. The problem is that Kubernetes is not designed to run experiments. Even if Kubernetes extensions exist for failure injection~\cite{chaomesh}, distributed monitoring~\cite{prometheus}, and workflow specifications~\cite{argo}, interconnecting these extensions involves significant domain knowledge and can be time-consuming. Additionally, it results in long and bloated specifications that make it difficult to trace the relationship between components and understand how they fit together.

We present \frisbee{}, a Kubernetes-native platform for the validation, verification, and testing of cloud-native applications. \frisbee{} provides the necessary abstractions to help developers focus on the test-case, rather than focusing on the technical functions of the testing mechanism. Tests are uniform, repeatable, and able to run as part of CI/CD workflows. Basic performance `smoke tests' can run as part of every build, while more extensive tests can run during acceptance testing phases.  \frisbee{} abstracts the testing process into four core components:
\begin{itemize}
    \item An integrated architecture for running application benchmarks, injecting faults, observing internal system interactions under evaluation (SUT), validating these interactions via test cases, and visualizing valid and invalid behaviors.

    \item Templates to describe the application's services, relationships between services, dependencies, requirements, and capabilities of each service. The service templates declare reusable types and use them to construct the structure of the application. 
    
    \item A distributed notification system that provides insights from different views (e.g., global, client, data servers), helping developers to respect logical dependencies, express different types of violations (e.g., data loss, availability), and reason for bugs closer to their source. 
    
    \item A declarative, human-readable language to create scenarios that simulate real activity and validate the application's behavior, both under normal and unexpected conditions.
\end{itemize}

Our main contribution is establishing the basis of a platform to carry out systematic testing of cloud-native applications over Kubernetes. While the various aspects (e.g., elasticity, availability, performance) can be tested individually in Kubernetes, this requires manual scripting and choosing from a wide ecosystem of Kubernetes extensions. Instead, \frisbee{} aims to be the first platform that provides a consistent way to design and run various classes of experiments over Kubernetes. To showcase its practicality, we use \frisbee{} for testing i) uncertainties at the level of application (e.g., dynamically changing request patterns), ii) failures at the level of infrastructure (e.g., crashes, network partitions), and iii) facts that drive deployment decision (e.g., saturation points). 

The rest of this paper is organized as follows: Section~\ref{sec:related} elaborates on related work and the advancements made by \frisbee{}, Section~\ref{sec:design} gives an overview of the \frisbee{} design. Section~\ref{sec:language} deepens into the language's abstractions. Section~\ref{sec:evaluation} presents a set of experiments that demonstrate the main aspects of the proposed language. Section~\ref{sec:conclusion} concludes the paper, and Section~\ref{sec:future} state our vision for the future.

\section{Related Work}
\label{sec:related}
To better understand the practicality of \frisbee{}, we compare it with five research directions that we regard as representative of its abilities: testing in the cloud, testing of the cloud, testing of the cloud in the cloud, chaos engineering, and testing specification.

\paragraph{Testing in the Cloud}
Testing in the cloud (TiC) refers to software testing performed by leveraging cloud technologies and resources to validate non-cloud software/applications, such as mobile or web environments. TiC overcomes the limits of traditional testing approaches, allowing testers to address testing objectives before considered infeasible. Among them is the possibility of performing massive combinatorial testing or evaluating attributes such as scalability, elasticity, and reliability by scaling up and down resources on-demand~\cite{gupta2011diecast}. However, TiC-based experiments require orchestration to be done manually via the vendor's API. A level above, Kubernetes-based designs present an advancement over the current state-of-the-art TiC, allowing engineers to write tests without having to consider the underlying infrastructure or the vendor's API~\cite{abedi2017conducting,cavalli2015survey}. With Kubernetes, experiments designed on a single workstation environment (e.g., MicroK8s) can be seamlessly executed on-premise (e.g., bare-metal K8s), on the Cloud via Kubernetes-as-a-service (e.g., Azure Kubernetes).

\paragraph{Testing of the Cloud}
Testing of the cloud (ToC) refers to validating the quality (functional and non-functional properties) of applications and systems deployed in the cloud. Simulators like CloudSim, and its derivatives~\cite{DBLP:journals/corr/GuptaDGB16,jha2020iotsim,sonmez2018edgecloudsim}, have been proposed as a cost-effective way to evaluate placement strategies and configurations for Cloud, Edge, and IoT environments. Though reproducible and convenient, simulators study behaviors expressed as mathematical models with certain assumptions, and therefore are inappropriate for testing real systems. Emulators incur similar infrastructure costs with simulators while achieving better realism since they run real codes. There are mainly two types of emulators: (i) Full-system emulators which use VMs as hosts to emulate heterogeneous nodes with various OSes ~\cite{gupta2011diecast, vishwanath2009modelnet}, and (ii) Container-based emulators leveraging containers that share a single OS kernel to provide better scalability, required to emulate large-scale networks~\cite{zeng2019emuedge, coutinho2018fogbed}. These tools are consumed with preparing the testbed rather than testing the application that runs on it. Approaches like MockFog~\cite{hasenburg2019mockfog}, Fogify~\cite{symeonides2020fogify}, and IOTier~\cite{nikolaidis2021iotier} have implemented ad-hoc alterations on a running testbed. However, these approaches are based on Docker Swarm, which does not support the sidecar pattern or automated control loops. Thus, that requires manual orchestration via the exposed APIs, and containers with monitoring tools pre-installed.

\paragraph{Testing of the Cloud in the Cloud}
Testing of the cloud in the cloud (ToiC) fills the intersection area between TiC and ToC: applications deployed in the cloud are tested by leveraging cloud platforms. The most related research direction in this category is Cloud-based Testing-as-a-Service Infrastructures. In general, tools in this direction provide a declarative environment for describing and executing automated performance tests. The focus can be on describing the deployment process and resource requirements~\cite{6825691}, describing elasticity when testing workflows with sequences of resource variation~\cite{10.1145/3023147.3023149}, automating the pipelines for deployment and monitoring~\cite{6427555,6676706, 7557407}, or abstracting and testing network configurations~\cite{7421896}. However, none of these works account for failure modes, and they are unsuitable for resiliency or availability tests. That is not surprising given that most Cloud providers did not provide capabilities for failure simulation until very recently. 

\paragraph{Chaos Engineering}
Chaos Engineering~\cite{basiri2016chaos} is now accepted as a fundamental procedure for ensuring that today's frequently changing and highly complex systems are achieving the resilience required. Chaos Engineering is the practice of injecting faults into a production system so that unanticipated weaknesses can be discovered and corrected before causing user issues. Systemic weaknesses could take the form of: improper fallback settings when a service is unavailable; retry storms from improperly tuned timeouts; outages when a downstream dependency receives too much traffic; cascading failures when a single point of failure crashes~\cite{haeberlen2005glacier}; etc. Chaos Engineering tools such as Chaos-Mesh~\cite{chaomesh} are excellent for injecting various faults into the system, but they do not provide the means for assessing the impact of a fault. This task is usually left to third-party tools. Services meshes, like Istio~\cite{istio}, provide both fault injection and observability of the fault. Their scope, however, is limited to the network level. In addition to the network, \frisbee{} operates at the application level as well. This involves a broader range of failures (e.g., kills, partitions, I/O) and application-level monitoring (e.g., number of ingested keys, replication errors, etc.).

\paragraph{Testing Specification}
A critical aspect of \frisbee{} is to help cloud system developers design and run test-cases more systematically than manual approaches. That requires abstracting the details of manipulating the state of an application under test and verifying the final application state against an expected state. Declarative testing is a test design paradigm that expresses tests in terms of data instead of procedural code. In this style, the testing rules (context, inputs, and expected outputs, defined by the user) are separate from the test automation code (implemented by the framework). However, existing frameworks are largely ineffective for Cloud-native applications. Firstly, they require access to source code~\cite{10.1145/1512762.1512764, 266952}. In contrast, constituent microservices of cloud-native applications are commonly delivered as containers. Second, framework are usually language-specific~\cite{269028, ginko, 6693140}. In contrast, microservices may be written in different languages, thus not always following a standard design pattern or sharing a common framework. Finally, although Kubernetes workflow engines, like Argo~\cite{argo}, facilitate the implementation of multi-stage experiments, these engines cannot simulate system failures or control the network nondeterminism; hence, it is difficult to write a test case to verify that a program is correct in the face of these conditions. To the best of our knowledge, \frisbee{} provides the first declarative language for testing Cloud-native applications, and the first testing framework that integrates natively to the Kubernetes ecosystem. 

\section{Design Overview}
\label{sec:design}
\begin{figure}[htbp]
    \centering
    \includegraphics[page=1, width=\columnwidth]{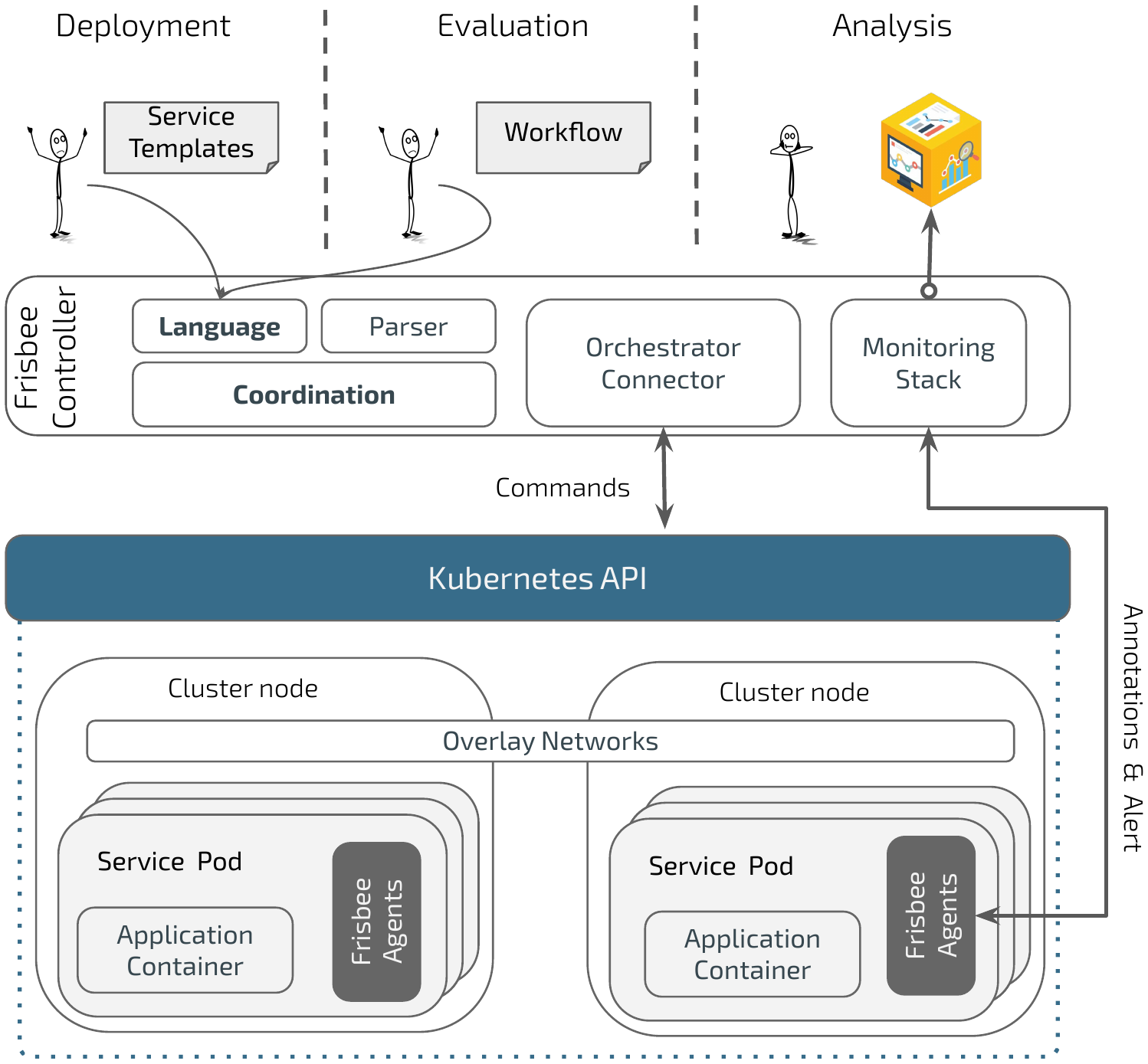}
    \caption{Frisbee architecture. Given a template describing the system under test and a workflow describing the experiment, \frisbee{} interfaces with Kubernetes to run the experiment. Notice the bidirectional feedback loop between Grafana and the controller(s).}
    \label{fig:architecture}
\end{figure}

Figure~\ref{fig:architecture} depicts a high-level overview of the \frisbee{} architecture. Adhering to the well-established Operator pattern of Kubernetes~\cite{operator}, it consists of  Custom Resource Definitions (CRDs) and a set of controllers integrating natively with Kubernetes services and APIs. A CRD  defines a new, unique object Kind in the cluster, listing out all of the configurations available to users of that object, and describing the desired state of that object. The controller is a control loop that watches the object’s state, making or requesting changes when needed to move the current state closer to the desired state. 

The experiment starts with specifying the CRDs in a YAML file. The most important ones include Service Templates and experimental Workflow. Templates describe the system under evaluation (SUT), including service definitions, such as Redis primaries or backups; logical dependencies between services, such as backup X depends on primary Y;
preferences on where services should be scheduled, such as on nodes with SSDs; and
resource constraints. The workflow is a graph that drives the evaluation process. The nodes represent actions that change the execution environment, and the edges represent assertions for the expected system behavior upon these actions. These assertions enable tests to run until some percentage of SLA violations occur, such as percentile latency exceeding a set limit, or a number of services have failed.

When the experiment is ready for deployment, the user passes the YAML to the workflow controller via the Kubernetes command line. The workflow controller is a global tester with total control over the SUT. It traverses through the graph and executes a specific set of operations based on the graph and the workflow’s current context. The context is driven by \frisbee{} events, which can be fired based on elapsed time, object state, or performance metrics. Since there is significant orchestration and sequencing involved, the workflow controller does not perform the actions itself but rather delegates them to other controllers to orchestrate their part. Note that each controller corresponds to a specific CRD so that the domain of each controller’s responsibility is clear. These sub-controllers may be native to Kubernetes (such as Pods or Deployments), belonging to the \frisbee{} ecosystem (such as the logical group), or be third-party (such as Chaos-Mesh). Sub-controllers translate higher-level actions into low-level operations, such as provisioning the requested resources, deploying the services, and disseminating policies to \frisbee{} agents at the execution layer. \frisbee{} uses sidecar agents to interact with application containers~\cite{burns2018designing,196346}, in a non-intrusive manner. Agents exist for resource and network throttling, monitoring shared resources, and suspending the container’s execution by preventing access to any resource.

In order to assure the correct assessment of a test, \frisbee{} gathers actionable data that provide not only the time an error or issue appears but, more importantly, its context. For example, a failure injection or a crashed node could explain a drop in requests. \frisbee{} follows a 3-factor approach to achieve observability: out-of-the-box monitoring, enrichment with control events, and automated alerts.

For the monitoring, \frisbee{} integrates several well-established tools in its stack: Telegraf for capturing performance metrics from the distributed services, Prometheus~\cite{prometheus} for gathering metrics into a centralized time-series database optimized for time-range queries, and Grafana~\cite{grafana} for real-time visualization, querying of metrics, and alerts. Every experiment runs in an isolated stack to prevent interference when running multiple experiments in parallel.

\frisbee{} also creates a feedback loop between Grafana and the controller. The feedback is bidirectional and serves a double purpose. First, when the controller performs an action, it pushes a descriptive annotation to Grafana. This annotation provides the context of the performance metrics to visually correlate the observed behavior with a root event. Second, when Grafana’s analytics methods fire an alert, this alert is propagated back to the controller, which decides what to do with it. For example, spin up a new server if the tail latency exceeds a given threshold or terminate the workflow indicating that the test has failed. 

Finally, \frisbee{} automates a critical step that is largely overlooked when testing a new system: clean-up. This includes removing generated artifacts (e.g.,  Kubernetes objects or data on remote hosts) and recovering from any injected faults (e.g., network partitions). Without consistent and diligent clean-up, the worst-case scenario is that changes in the environment can cause test `flakiness’. These differences may be so small that engineers will not spot them and will deliver false test results.

\section{The \frisbee{} Language}
\label{sec:language}
\begin{figure}[htbp]
    \centering
    \includegraphics[page=2, width=\columnwidth]{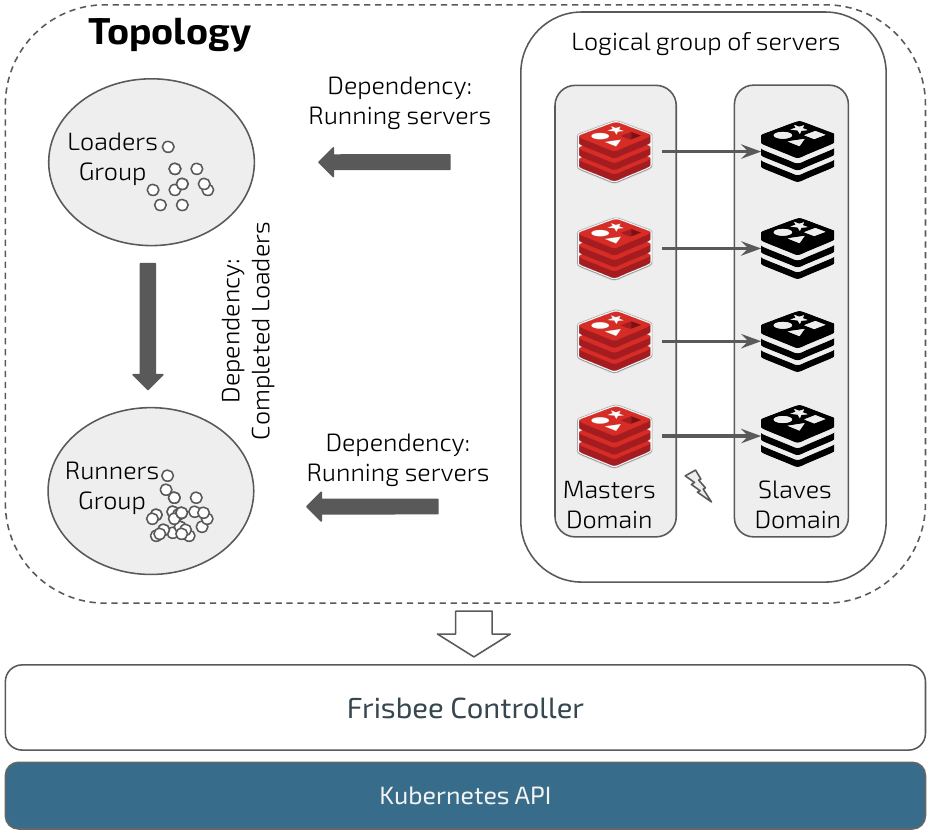}
    \caption{Redis deployment example. Masters must be running before starting the load phase. The Load phase must complete before starting the Run phase. Replicas must run on a different failure domain than Masters.}
    \label{fig:example}
\end{figure}

To better understand the complexity \frisbee{} aims to address, consider the setup of Figure~\ref{fig:example}. Let us assume a cluster of 4 Redis~\cite{redis} masters replicating to 4 Redis slaves, with 10 parallel YCSB client (loaders) inserting keys and 5 YCSB~\cite{cooper2010benchmarking} clients (runners) executing mixed queries against the inserted keys. Redis masters serve both read and write requests, while slaves serve only read requests. Then, we want to spread Redis masters and Redis slaves to different availability zones to reduce correlated failures. The experiment consists of three distinct phases; Setup, Load, Run. These phases have logical and timely dependencies that must be respected. The key ingestion does not start unless the servers (both masters and slaves) are up and running. Similarly, the queries must be postponed until the distributed loaders have completed ingesting the keys. 

\begin{code}
\begin{minted}{yaml} 
Kind: Template
metadata: {name: redis}
spec:
 "master":
   inputs:
    parameters:
     port: "6379"
   spec: |
    resources: {cpu: 4, mem: 4GB}
    agents:
     telemetry: [sysmon/container, redismon/telegraf]
    container:
     name: app
     image: redis
     ports:
      - name: to-clients
        containerPort: {{.inputs.parameters.port}}
      - name: to-cluster
        containerPort: {{add .inputs.parameters.port 10000}}
     command: ....
\end{minted}
\captionof{listing}{Snippet of Service Templates. The embedded scripts reduces the complexity of defining a service, exposing only parameters that matter to the user. Parameters are evaluated at runtime.}
\label{code:service}
\end{code}

\subsection{Service Templates}
Kubernetes users often find themselves copying YAML specifications and editing them to their needs. This approach makes it difficult to go back to the source material and incorporate any improvements made to it. The template allows users to create a library of frequently-used specifications and reuse them throughout the definition of the experiment. These specifications define the container image to use,  the commands to execute inside containers, the ports where the containerized application is listening for requests, CPU and memory limits and reservations, and environment variables used as inputs to the containers.

Templates define minimally constraining skeletons, leaving a bunch of strategically predefined blanks where dynamic data will be injected to create multiple variants of the specification using different sets of parameters (Code~\ref{code:service}). In contrast to other templating engines that are evaluated at deployment time (e.g., Helm), \frisbee{} templates are evaluated in runtime to be usable by calling objects, such as Workflows. Users may still develop their Services using familiar Kubernetes constructs with the added functionality of \frisbee{} not affecting portability. The means that a \frisbee{} enhanced description will run in any Kubernetes environment without any alternations; however, users will lose the convenience offered by \frisbee{}.

\paragraph{Embedded Scripting}
Without parameters, templates provide reasonable defaults for the described services. With given parameters, \frisbee{} executes any embedded scripting in the template (denoted as \{\{...\}\}) and replaces parameters with desired values — to generate the configuration. Embedded scripting allows us to reduce user exposure to configuration parameters. For example, Every Redis Cluster node requires two TCP connections open. The standard Redis TCP port used to serve clients, for example, 6379, plus the port obtained by adding 10000 to the data port, so 16379 in the example. For port forwarding purposes, these ports must also be declared in the container specification. Embedded scripts can automate this remapping and reduce the complexity of defining a service, exposing only parameters that matter to the user.

\paragraph{Resource Throttling}
\frisbee{} allows users to specify the amount of resources that each service will use. Limits/quotas can be specified for memory, CPU, disk I/O, and network I/O. Memory limits prevent the container from using more than a given amount of user and system memory. For instance, if a container is allocated 4GB, \frisbee{} ensures that the specified limits are not exceeded. For the allocation of CPU cycles, we use the CPU units of Kubernetes. One unit is equivalent to 1 vCPU/core for Cloud providers and one hyperthread on bare-metal processors. Fractional (less than 1) requests are allowed. A request of 0.5 CPU is guaranteed half as much CPU as one that asks for 1 CPU. If an offending instance exceeds these limits, Kubernetes disallows it from using more of the specified resource.



\paragraph{Observability Packages}
Despite the numerous telemetry agents and Grafana dashboards, only a few are interoperable. Templates enable us to create observability packages that bundle a particular service with a telemetry agent and a set of pre-configured Grafana dashboards and alerts. Installation of these packages involves activity in two places: (i) in Grafana for configuring the dashboards and alerts, (ii) in the pod for deploying the telemetry agent. When someone asks for a service, the template controller automatically deploys the corresponding dashboards into Grafana. It is up to the caller (e.g., Workflow controller) to deploy the telemetry agent alongside the requests service.

\subsection{Distributed notifications}
Reasoning about the system's behavior in dynamically evolving environments is challenging in its own right. An innovative aspect of \frisbee{} is that it exposes the system state into top-level events that pass contextual properties between controllers, enabling transactionality and idempotency of operations. Events can be used to schedule actions, handle logical dependencies between steps, or test oracles to decide if the test has passed or failed. We support four types of events: 

\paragraph{Time events:} are fired after an elapsed time measured by the controller. Time events have three different formats that underpin their function. The first format (e.g., "10s"), waits for the given duration to elapse, and then it is fired once. The second format (e.g., "3m@5") is similar but repeatable regarding the given number. For example, this event can be used to create a new YCSB runner every 3 minutes, with 5 runners added in total. The third format is compatible with cron (e.g., 10 14 * * 1) and can be used to run periodically at fixed times, dates, or intervals. 

\paragraph{State events:} transcribe to the state of Kubernetes objects~\cite{pod-lifecycle}. They are automatically created when objects have state changes, errors,  or other messages that should be broadcast to the system. The \frisbee{} controllers typically run a watchdog that listens for events on the objects they own. For example, the Workflow controller listens for events created by sub-controllers, like the group controller. In its turn, the group controller listens for events on the services it owns. 

\paragraph{Performance events:} exploit Grafana's ability to perform statistical analysis on key performance indicators (KPIs) and fire an alert if the outcome matches a given rule. Performance indicators can be the response time, system throughput, utilization of system resources, execution time, and error rate. Rules may be as simple as `response\_time > 200ms', or more complex and involve the rate of change of a given metric. For example, assuming that A represents the incoming traffic, the expression `WHEN percent\_diff () OF query (A, 5m, now-15m) IS ABOVE 20' is roughly translated as `raise an alert if the traffic from now is more than 20\% of some reference ago (5min, 1min, etc.)'. 
 
\textbf{Annotation events:} are used to pass contextual information between the controllers. For example, we use these events to ask a controller to gracefully terminate one of its services or inform the owner that a particular service will be killed through a Chaos experiment. Although annotations are eventually available to the user via the object's metadata, the underlying synchronization mechanism remains internal into \frisbee{}.


\begin{table}[htp]
\centering
\begin{adjustbox}{max width=\columnwidth}

\begin{tabular}[b]{|l l|} 
 \hline
 Normal & Description \\ 
 \hline\hline
 Create & Creates a new element \\ 
 Update & Updates a running element \\
 Stop & Gracefully terminates a running element \\
 Pause & Suspend the execution of an element \\
 Resume & Resumes the execution of an element \\ 
 Teardown & Gracefully terminate experiment \\
 Destroy & Ungracefully terminate experiment \\
 \hline\hline
 Chaos & Description \\ 
 \hline\hline
 Kill & Forcibly terminates a running element  \\
 Partition & Simulates network partition.                   \\
 Netem & Simulates Network conditions (delay, duplicates)   \\
 IO & Simulate I/O faults (I/O delay, read/write errors)    \\
 Revoke & Cancel the previous failures \\ 
 \hline\hline
 Control & Description \\ 
 \hline\hline
  Depends & Wait for an event before an action  \\
  Oracle & Validate the system state after an action  \\
 \hline
\end{tabular}

\end{adjustbox}
\caption{List of \frisbee{} Actions. Normal actions simulate uncertainties that occur naturally to an Internet-facing system. Chaos actions inject faults to simulate abnormal operating conditions. Control specify the dependencies between actions.}
\label{table:actions}
\end{table}

\subsection{Workflow Specification}
The workflow is the end-to-end description of a \frisbee{} experiment. It serves two important functions:
\begin{enumerate*}[(i)]
    \item it defines a set of actions to be executed, and
    \item it defines the desired state of the system after each action.
\end{enumerate*}

In this view, the workflow is not a pre-determined configuration but rather a `live' object that traces the execution of the deployed application and schedules the execution of actions. 
However, since there is significant orchestration and sequencing involved, the workflow controller does not perform the actions itself but rather delegates them to other controllers to orchestrate their part. Their general format is \textit{action name: dependencies: action type: test oracle}. The appropriate controller is selected based on the type of action.

\paragraph{Actions}
Actions can be thought of as `functions' that define instructions to be executed. Table~\ref{table:actions} depicts the supported actions and Code~\ref{code:scenario} shows how to use them. 

\textbf{Normal conditions:} These actions simulate uncertainties that occur naturally to an Internet-facing system, such as joining or leaving services. We discuss this controller further in the following Subsection. 

\textbf{Abnormal conditions:} These actions inject faults to simulate abnormal operating conditions, such as network partitions or DNS errors. A fault causing a failure may be permanent or transient. For example, killing a service is a permanent fault. Oppositely, a network partition is transient. Partitioned nodes will experience a connection outage for as long as the fault remains active, but they become operational once the fault is revoked. Faults can be revoked either by duration or manually via the Revoke() action. Thus, we can support repeatable and complex failure patterns (e.g., inject a partition fault every 2 minutes to simulate intermittent connectivity).

\textbf{Wrappers:}
In many cases, we want to manage Kubernetes resources from the \frisbee{} workflows. This is common when the controller of the target resource is based on a different Kubernetes version than the \frisbee{} controller. In this case, the two controllers cannot interact directly but only by creating, deleting, or updating the CRD of the target resource. Practically, the wrappers enable \frisbee{} to interact with any third-party Kubernetes extension.

\paragraph{Dependencies}
For modeling multi-stage experiments, users can define the workflow as a directed-acyclic graph (DAG) by specifying the dependencies of each action. Dependencies are based on fields exposed by Time, State, and Performance events. However, there are cases in which we want to evaluate numerous events to see if certain conditions are met. For instance, to start queries only when all servers are running and loaders are complete. To do that, we adhere to the syntax of govaluate, which covers all the regular C-like expressions (e.g., logical/arithmetic/string comparators). As syntactic sugar, we also support YAML-like expressions that denote the logical AND of test oracles of the named objects, as shown in Code~\ref{code:scenario}.

\paragraph{Test Oracle}
The oracle is a mechanism for determining whether a test has passed or failed. The use of oracles involves comparing the output(s) of the system under test, for a given test-case input, to the output(s) that the oracle determines that product should have. In general, the syntax of an oracle is similar to that of dependencies. However, they are semantically different. The dependencies are "pre-action" conditions, whereas the oracles are the "post-action" conditions denoting expected behavior. 
Another difference is that dependencies conceptually belong to the level of a workflow and can only be set within its scope. In contrast, oracles are specific to objects and can be set within them. As will become apparent in the next Subsection, this design decision allows us to define object-specific conditions, such as the number of failures a group can tolerate.

\paragraph{Expected versus Unexpected Failures}
A service may fail due to bugs in the code or due to faults injected within the context of a Chaos experiment. In the first case, the failure is unexpected, and the test should fail immediately. In the second case, failure is expected, and we should let the test continue. Thus, we need to identify the failure type and distinguish the expected from the unexpected failures. The basis of our mechanism is that most fault injection tools operate at the level of a Pod, and therefore the affected services are known in advance. Whenever the \frisbee{} workflow controller encounters a fault action (e.g., kill, partition), it informs (via annotations) the service owner (i.e, logical group) before it applies the action to the service. The user can incorporate this annotation into oracles, further restricting deviations from the expected behavior.

\paragraph{Termination \& Clean-up}
A Workflow consists of multiple objects with a parent-child relationship  between them, like the following:  $Workflow \rightarrow Create \rightarrow Group \rightarrow Services \rightarrow Pods$.  The root is the workflow, then actions are children to the root, the objects managed by \frisbee{} are children to the actions, and the leaves are pointers to objects managed by third-party controllers.  For example, the default Kubernetes controller manages Pods, and the Chaos-Mesh controller manages failures.
By traversing this dependency tree, the deletion propagates down to the children when the parent is deleted. Practically, this means that if we delete the workflow, all the remaining objects created throughout the experiment will be automatically garbage-collected by Kubernetes. The opposite direction, how a failure in the pod causes a workflow failure, depends on the test oracle of the group owning the failing pod.





\begin{code}
\begin{minted}{yaml} 
Kind: Workflow
metadata: {name: redis-failover}
spec:
 # Create a Master node
 - action: DistributedGroup
   name: masters
   distributedGroup:
    templateRef: redis/master
    instances: 1
 # Create a Slave node    
 - action: DistributedGroup
   name: slaves
   depends: { running: [ masters ] }   
   distributedGroup:
    templateRef: redis/slave
    inputs:
     - {master: .group.masters.any}
# Create a failover manager     
 - action: DistributedGroup
   name: sentinel
   depends: {running: [masters, slaves]}
   distributedGroup:
    templateRef: redis/sentinel
    inputs:
     - { master: .group.masters.all }
# Ingest keys in parallel into different ranges  
 - action: DistributedGroup
   name: loaders
   depends: {running: [masters, slaves, sentinel]}
   distributedGroup:
    templateRef: redis/loader
    inputs:
     - { server: .group.masters.any, offset: "0" }
     - { server: .group.masters.any, offset: "100000" }
     - { server: .group.masters.any, offset: "200000" }
# Run queries
 - action: DistributedGroup
   name: runners
   depends: {running: [masters, slaves], success: [loaders]}
   distributedGroup:
    templateRef: redis/runner
    instances: 5
    schedule: "@every 2m"
    inputs:
     - { server: .group.masters.any, workload: workloada}
\end{minted}
\captionof{listing}{Workflow DAG with dependencies, input method, macros, scheduling constraints. For brevity, we use the YAML syntax for dependencies and assume groups with default lifecycle.}
\label{code:scenario}
\end{code}

\subsection{Logical Groups}
The \textit{group} is an abstraction for running multiple services in a shared context. It oversees the creation and the management of the lifecycle of constituent services, thus saving users from the trouble of tracking individually failing services. 

\paragraph{Create services}
The group's specification consists of a reference to the service template and directives for the creation of service instances.  As shown in Code~\ref{code:scenario}, the are two ways to create a group of services. For identical services, it suffices to specify the number of desired instances. For services with distinct parameters, the user can specify a list of inputs that \frisbee{} will iterate over and use them for calling the template. Users may also specify the creation order. We support three creation models: (i) sequential: one service created immediately after the other, (ii), parallel where multiple services are created simultaneously, and (ii) scheduled: like sequential, but instead of being immediate, it is based on Time events. By spreading parameterized instances over time, e.g., create a new service every 2 minutes.,  we can create variable workloads and dynamically changing topologies while maintaining the shared execution context between instances. 

\paragraph{Dynamic membership}
For post-analysis purposes, grouped instances are placed in a hierarchical namespace prefixed by the group's name (e.g., Master-0, Master-1, Master-N). However, the dynamic membership can cause numerous issues, such as addressing a pending service or a removed service. For the consistent and scalable access of grouped services, we provide macros in the format `.{groupName}.{filter}'. The predicate `groupname' points to the desired group. The predicate filter selects a set of Services within the group. Available filters include first(), last(), oldest(), recent(), all(), any(), percent(). When \frisbee{} 's interpreter stumbles onto a macro, the  \frisbee{} controller scans its internal tables, keeping entries for the group, and selects the returned services according to the filter. For example, the macro `.masters.*' (alias to `.master.all`), will return all the services from the master group. The first() and last() select based on alphabetical order, oldest() and recent() select based on the creation timestamp, any() selects a random running service, and percent() selects a random subset of the running services. The percent is useful for causing correlated failures to services (e.g., kill 3\% of the services). 

This layer of indirection also facilitates the scalability of an experiment. With macros, a group can scale from 10 services to 100 services without changing the addressing semantics. Without macros, users had to hardcode and manage yet another 90 services in the experiment. Additionally, it allows the user to choose whether failing services will be addressable by macros or not.

\paragraph{Informed Placement}
Albeit operated in a shared context, services run in separate pods to achieve isolation between them. By default, Kubernetes assigns pods to nodes based on available resources. Nonetheless, there are some circumstances where users need to control the placement manually. Although Kubernetes allows users to set placement constraints via the node and affinity/anti-affinity selectors, these primitives are low-level and only applicable to the level of individual Pods. Groups provide the necessary level of abstraction to simplify the placement of services.  Their semantics can be briefly summarized as \textit{All services belonging to the same group are scheduled on the same set of physical nodes and networks, while preventing services belonging to a different group to run on the same set of resources}. 

When combined with labels on the physical nodes, groups can be used for scheduling services on those nodes whose labels are identical to the labels defined in the group description. For example, groups with SSD labels will be scheduled only on nodes with SSD storage attached. Alternatively, by creating two different groups, we can have a few nodes dedicated to compute-intensive services and others with enough CPU and RAM dedicated to memory-intensive Services. Thus we prevent undesired services from consuming resources dedicated to other services.

\paragraph{Lifecycle Management}
By default, the group follows a defined lifecycle. Starting in the \textit{Pending} phase, moving through \textit{Running} if at least one of its children is running, and then to \textit{Success} when all services have exited successfully, or to \textit{Failed} if any of the children have failed. Albeit reasonable in failure-free scenarios, this lifecycle may cause erroneous behaviors when used in conjunction with Chaos experiments. For example, to benchmark the recoverability of a system, we need to inject a fault and let the system under evaluation perform its recovery tasks (e.g., rebalance the load on the remaining nodes). However, if the service fails, then the group will fail, and the failure will be propagated back to the workflow (via the State events), which then will fail itself. Thus, a failed services will lead to the premature ending of the entire workflow, which is not the desired behavior. To handle this issue, we use a test oracle that drives the lifecycle of a group: 

\textbf{Fail:} indicates the conditions under which a group is considered to have failed. This way, we can explicitly refer to how many failures a group can tolerate before it fails itself. 

\textbf{Success:} indicate the conditions under which a group is considered as completed and can be garbage collected. This oracle makes it possible to create `fixed-percent' dependencies to satisfy the group dependency even if only a given percent of the services satisfy it. 

\textbf{Suspend:} indicate the conditions under which the group should suspend the execution of new services. As we will explain in the evaluation Section, we use this oracle for detecting saturation points.

\section{Evaluation}
\label{sec:evaluation}
To showcase the need for and the functionality of \frisbee{} we used the Redis key-value store to highlight the contributions of this work. At every experiment, we report the methodology, the lines it took to describe the respective workflow, the exercised components of \frisbee{}, and how they helped us to perform the test and reason about the system's behavior. 

For the evaluation, we used a Kubernetes cluster of 4 server-grade nodes. Each node has 24-cores (2 Intel Xeon E5-2630v3 CPUs), 128 GB of DDR3 ECC main memory, and 250 GB of locally attached NVMe storage. The machines are connected via 1Gbps links to the top-of-rack switch. Additionally to the server-grade nodes, we used a workstation with 8-cores (Intel i7-7700 CPU), 32 GB of DDR4 main memory, and 256 GB NVMe storage. For monitoring we used the bitnami containers Telegraf, Prometheus, and Grafana. For failure injection, we use Chaos-Mesh~\cite{chaomesh}. 

\subsection{Application: Load test}
This experiment verifies that Redis performs well when used by a normal number of concurrent users, with dynamically changing request patterns. This is needed since most benchmarks generate a single, stationary workload per run, which is not representative of cloud-native environments that run various mixes of I/O workloads. This experiment exercises the utility of: (i) logical dependencies, (ii) scheduling constraints, and  (iii) contextualized visualization. The workflow description is 66 lines. 

\paragraph{Methodology}
We divide the experiments into phases, and at each phase, we assemble a different combination of YCSB workloads together to form a synthetic workload. We have four phases: boot, load, scale-up, and scale-down. The boot phase represents the creation Redis server. The load phase has a WaitReady dependency on the boot phase, meaning that YCSB clients must wait for Redis servers to be up and running before starting inserting keys. The scale-up phase has a WaitExit dependency to load phase -- meaning that all loaders must complete (event 3) before starting to run queries (events 4-8). Runners in this phase are scheduled to run every two minutes and execute different YCSB workloads in the following order: [workloadA, workloadB, workloadC, workloadD, workloadA]. WorkloadA has a mix of 50\%/50\% reads and writes. WorkloadB has 95\%/5\% reads/writes mix. WorkloadC is 100\% reads. WorkloadD inserts new records and runs queries against the most recent keys. This allows us to run a given load over a long period of time, as a realistic test of a production environment, and have transient clients as a way to emulate `noisy neighbors'.

\begin{figure}[tbp]
    \centering
    \includegraphics[page=1, width=\columnwidth]{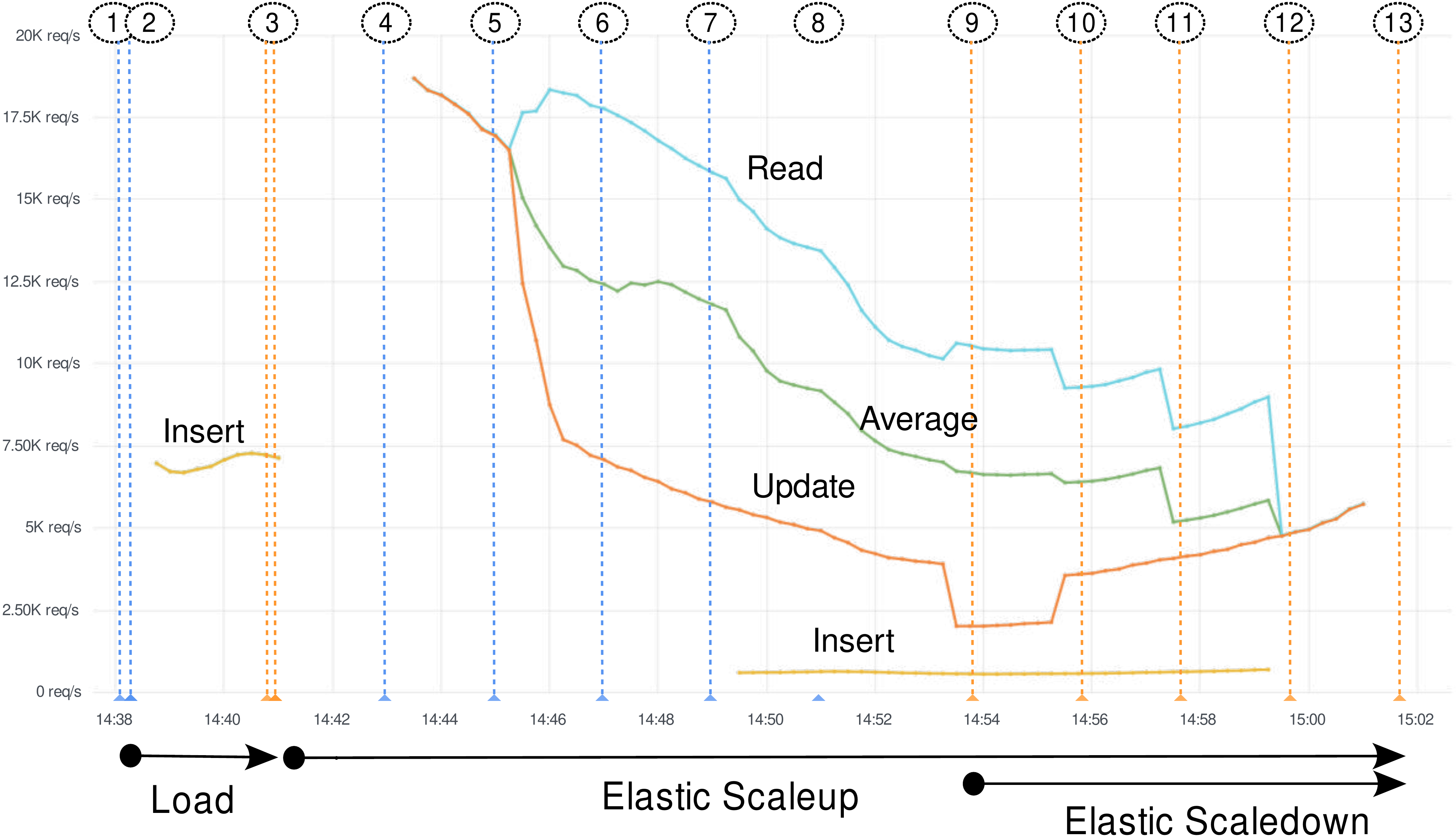}
    \qquad
    
    \begin{adjustbox}{max width=\columnwidth}
    \begin{tabular}[b]{l | l | l | l }
      \toprule
      Phase [Event]  & Depends &  Action & Description  \\\midrule
boot      [1]   & none               & Create(Redis, 1)        & Add 1 redis server                  \\
load      [2,3]   & WaitReady(boot)    & Create(Loader, 10)    & Add 10 parallel loaders             \\
scaleup   [4-8]   & WaitExit(scaleup)   & Create(Runner, 5, 2m) & Create 5 runners, 1 every 2 minutes \\
scaledown [9-13]  & WaitReady(scaleup)   & Stop(Runner, -, 2m)   & Stop all runners, 1 every 2 minutes \\
      \bottomrule
    \end{tabular}
    \end{adjustbox}
    
  \caption{Average YCSB throughput, with annotations for joining clients, and failure periods.}
\label{eval:elasticity}
\end{figure}

\paragraph{Identifying contention}
Figure~\ref{eval:elasticity} depicts how the average per-client queries per second (throughput) evolves through the experiment's lifetime. Every phase starts and ends with a unique event, depicted on the top. These events describe when services are created (blue line) or terminated (orange line). Because the timestamp is measured by request acknowledgment, there can be a skew of a few seconds between the annotated event and the observed behavior. 

Starting from Event 4, we can see the overlapping of Read and Update operations. At Event 5, when a new runner with workloadB has joined, there is a rapid drop in the average performance. Notice that performance effects are accumulative. On Event 4, there is a single runner. On Event 5, there are two runners and three runners on Event 6. Thus, the overall system bandwidth is shared between them. A key to understanding this behavior is that the reported throughput is the per-client average and not the system average. Later, on Event 7, the new runner, running workloadD, applies insert operations, reducing the average throughput. This drop is caused by contention in the top-lock of Redis, which produces interference between the update and insert operations.

\paragraph{Ensuring Logical Dependencies}
A more peculiar dependency is that between scale-down and scale-up. Without any dependency, the scale-down would start immediately after scale-up, causing created services to terminate immediately. For this reason, the scale-down must wait for the scale-up group (meaning all its services) to be up and running. Given the non-deterministic nature of Kubernetes, simplistic time-based approaches would likely lead to inconsistent results through different test iterations. With \frisbee{} state events, we can guarantee consistency in respecting the logical dependencies between the various stages of an experiment.

\begin{figure}
\subfloat[Aggregated]{\includegraphics[page=1, width=.9\columnwidth]{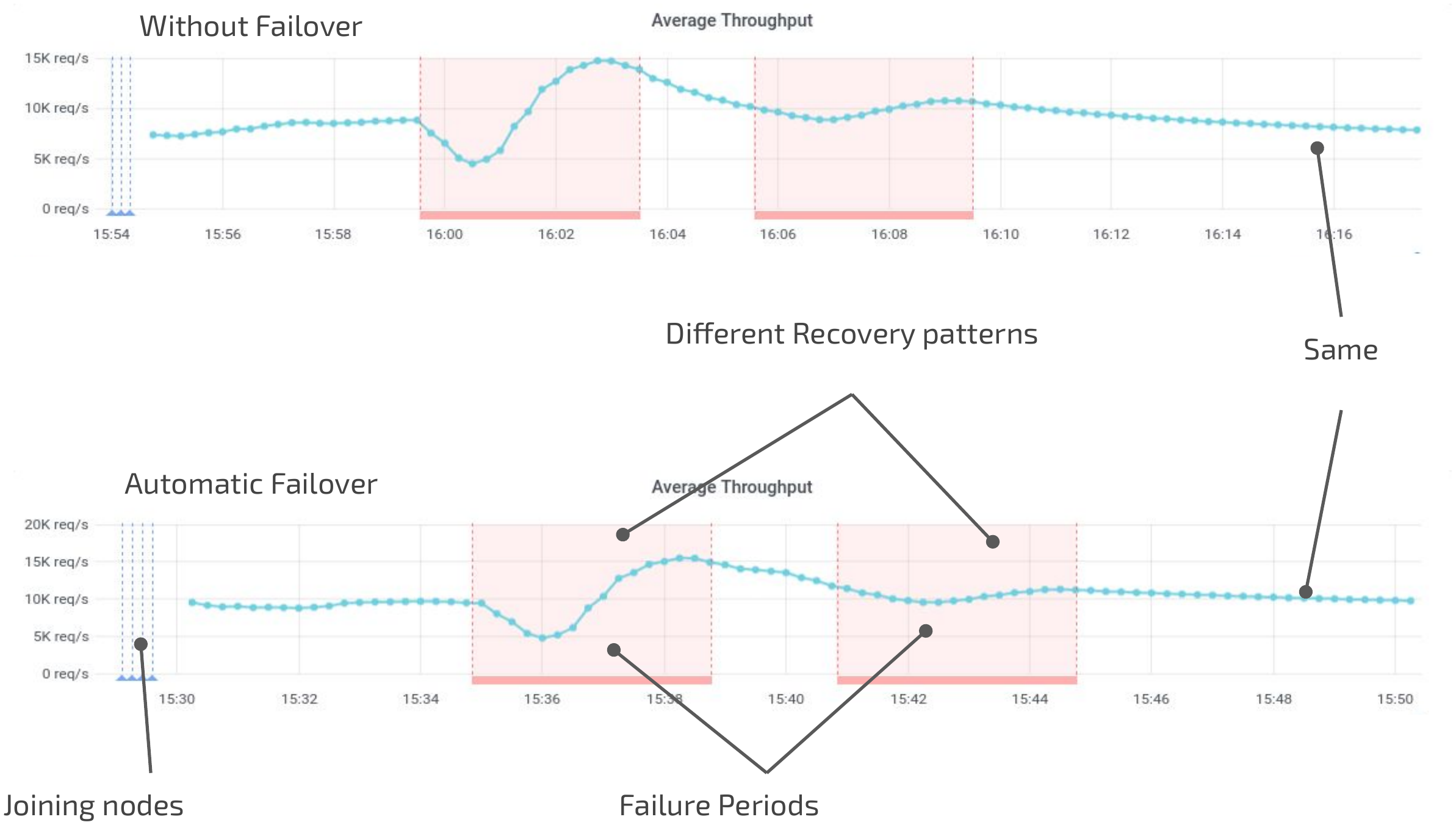}} 
\\ 
\subfloat[Per-operation]{\includegraphics[page=1,width=.9\columnwidth]{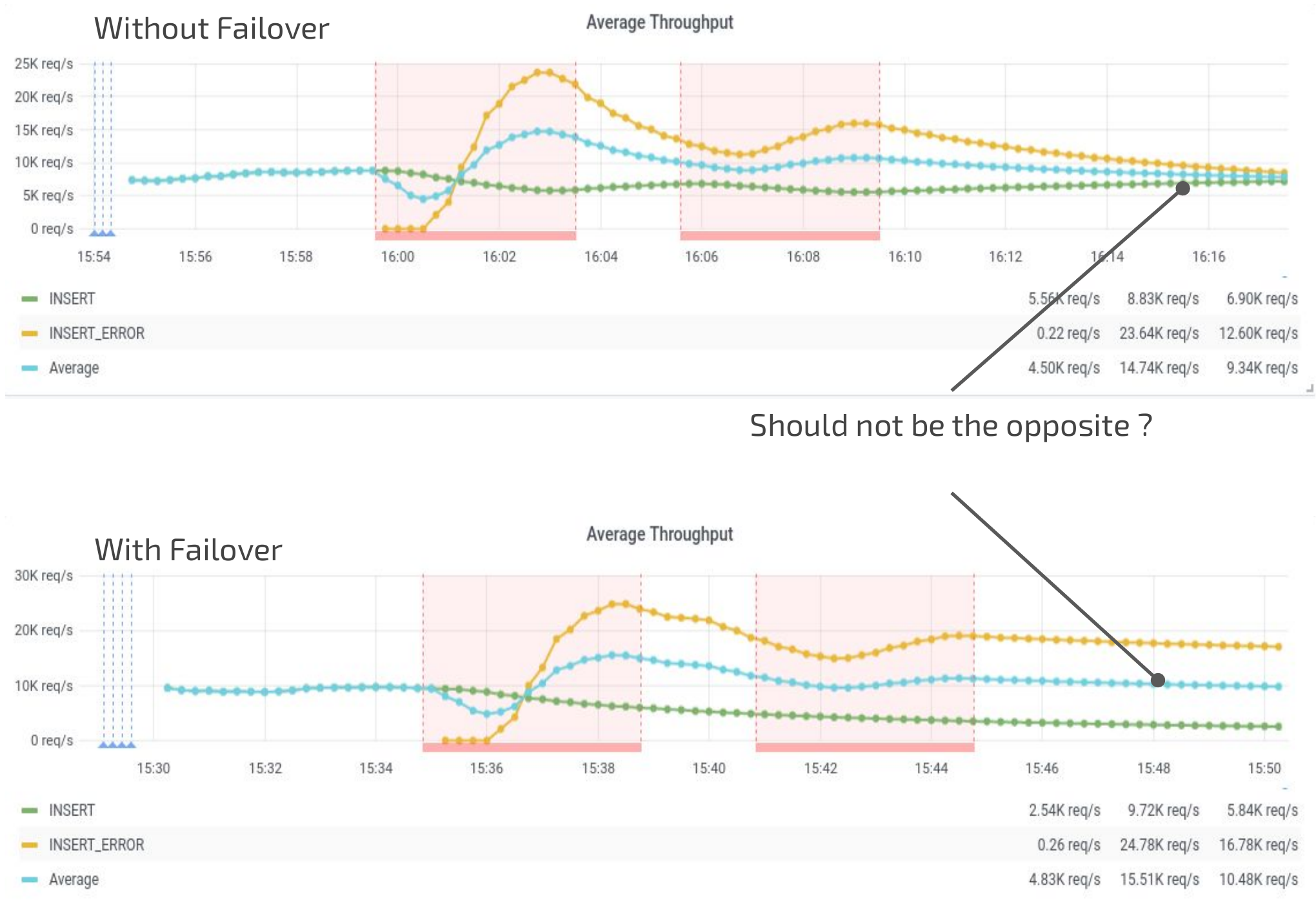}}
\\
\caption{Average YCSB throughput, with annotations for joining clients, and failure periods.}
\label{eval:recovery:client}
\end{figure}

\begin{figure}
\subfloat[Master view]{\includegraphics[page=3,width=.9\columnwidth]{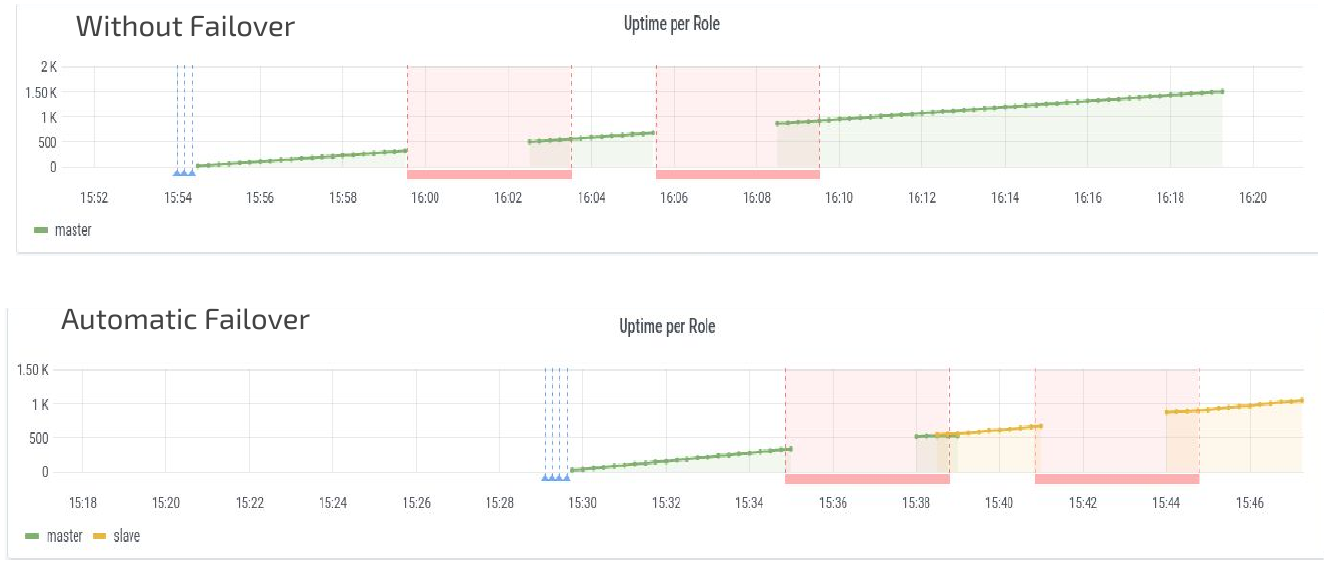}} 
\\ 
\subfloat[Slave view]{\includegraphics[page=4,width=.9\columnwidth]{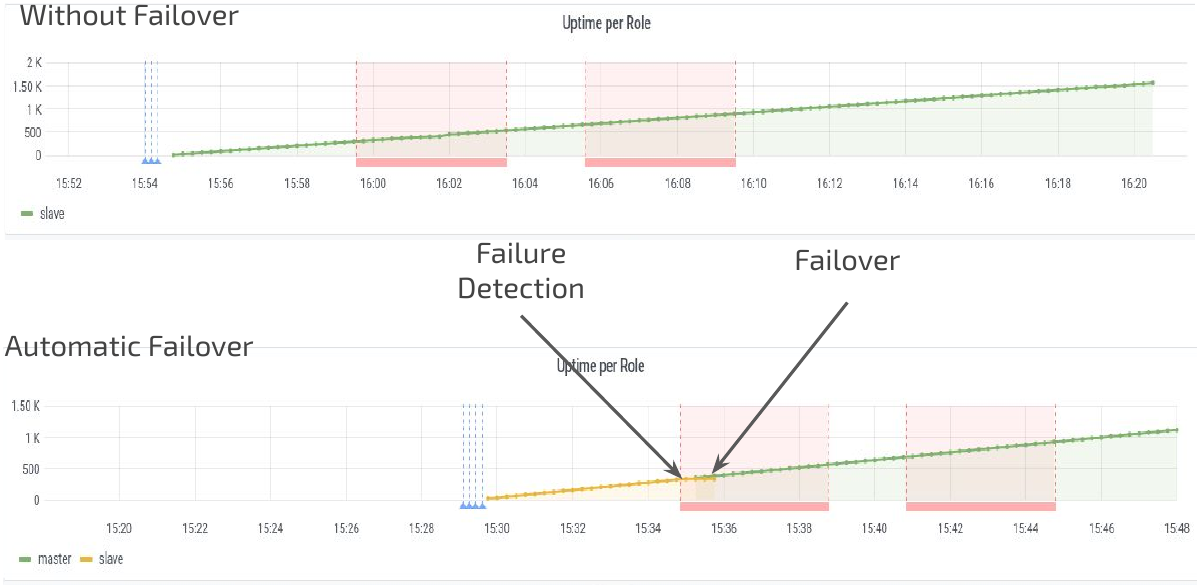}}
\\
\caption{Evolution of Redis servers, with and without failover. Without failover, the Master and Slave retain the roles for the whole experiment. With automatic failover, they swap roles.}
\label{eval:recovery:server}
\end{figure}

\subsection{Infrastucture: Failover test}
With failures being unavoidable, a system's ability to recover from failures quickly is a critical factor in the overall availability of the system~\cite{10.1145/3447851.3458738}. This experiment studies the Redis ability to replace a failed component under a network partition, with minimal service disruption. This experiment exercises the utility of: (i) failure injection (ii) failure retraction (iii) placement constraints, and (iv) application-level monitoring from different views (client, master, slave). The workflow description is 72 lines. 

\paragraph{Methodology}
We take a simple topology with two replicated Redis servers and one YCSB client placed on different physical nodes. We assume an intermittent partition failure that makes the Redis master appear and disappear periodically. We study two cases: with and without automatic failover. We use the Redis Sentinel for the automatic failover, which will promote a slave node to a master in case the master fails. Figure~\ref{eval:recovery:client} provides information about the time that nodes were created, the failure periods, and the average client queries per second (throughput) during those periods. There are two important observations. 

\paragraph{Effects of Aggregated Values}
The first observation is that the recovery behavior appears to be different between subsequent failures. Whereas there is a significant fluctuation in performance in the first failure, the system appears relatively stable in the second failure. A deeper investigation reveals that YCSB uses the same format for successful and failed operations. As a result, the failed operations also contribute to calculating the average. With errors starting low and rapidly increasing in the first failure, the average appears to fluctuate significantly in both directions. In the second failure, the failed operations are already high, thus impacting the average less. An important takeaway is that YCSB keeps sending requests continuously, regardless of the outcome of past requests. However, failed requests are acknowledged faster than successful requests (fewer software layers to traverse). Thereby, the apparent throughput of failed requests can be significantly higher than the throughput of successful requests - albeit erroneous. Without \frisbee{} automation for collecting distributed metrics and putting them into analytic dashboards, we would have likely overlooked this point, leading us to unfair comparisons and invalid conclusions.

\paragraph{Failover Effectiveness}
The second observation is that without failover, the system appears to recover after the failures are retracted. The failed requests are reduced, and the successful requests are increased. With automatic failover, the failed operations remain high, whereas the successful operations are decreasing. This behavior is the exact opposite of what someone would expect. To understand what was happening, we had to study the failover from the side of the servers. Figure~\ref{eval:recovery:server} shows the role of each node over time.  

Without failover, the master and slave nodes retain their roles for the whole experiment. The gaps appear because the network partition on the master node affects connectivity with all the services, including the monitoring tools. After the partition failure is retracted, the master returns to service. For this reason, the system appears as if it has self-recovered. With failover, studied from the perspective of the master node, we see that the master and slave swapped roles sometime during the failure. However, the connectivity gaps limit our knowledge about what exactly was happening during the failure. The visualizations, from the perspective of slave nodes, solve this deficit. By querying the slave metrics, we can measure Redis' resiliency in terms of 
i) the magnitude of deviation from the nominative performance levels during the partition,
ii) the sensitivity in detecting the failure,
iii) the time it takes to regain nominal performance levels, and, 
iv) the completeness of recovery in a finite amount of time.

\paragraph{Implicit dependencies}
Despite the validation that master and slaves switch roles during the partition, further investigation was needed to identify why the client behaved as if failover did not happen. A thorough exploration of the Redis documentation revealed that for failover to work, the Redis clients must have explicit support for the Sentinel protocol. Unfortunately, the Redis client of the YCSB benchmark does not support it. Although Sentinel successfully swapped the two servers during the failover, the client was not updated and kept sending requests to the master node experiencing the failure. However, since the master was downgraded to slave, it could no longer serve insert requests, even when the connection was restored. Regardless of the specific reason, this experience highlights the need to integrate failures in the systems evaluation methodology. 

\begin{figure}[htb]
\subfloat[Server Operations/second]{\includegraphics[page=1,width=\columnwidth]{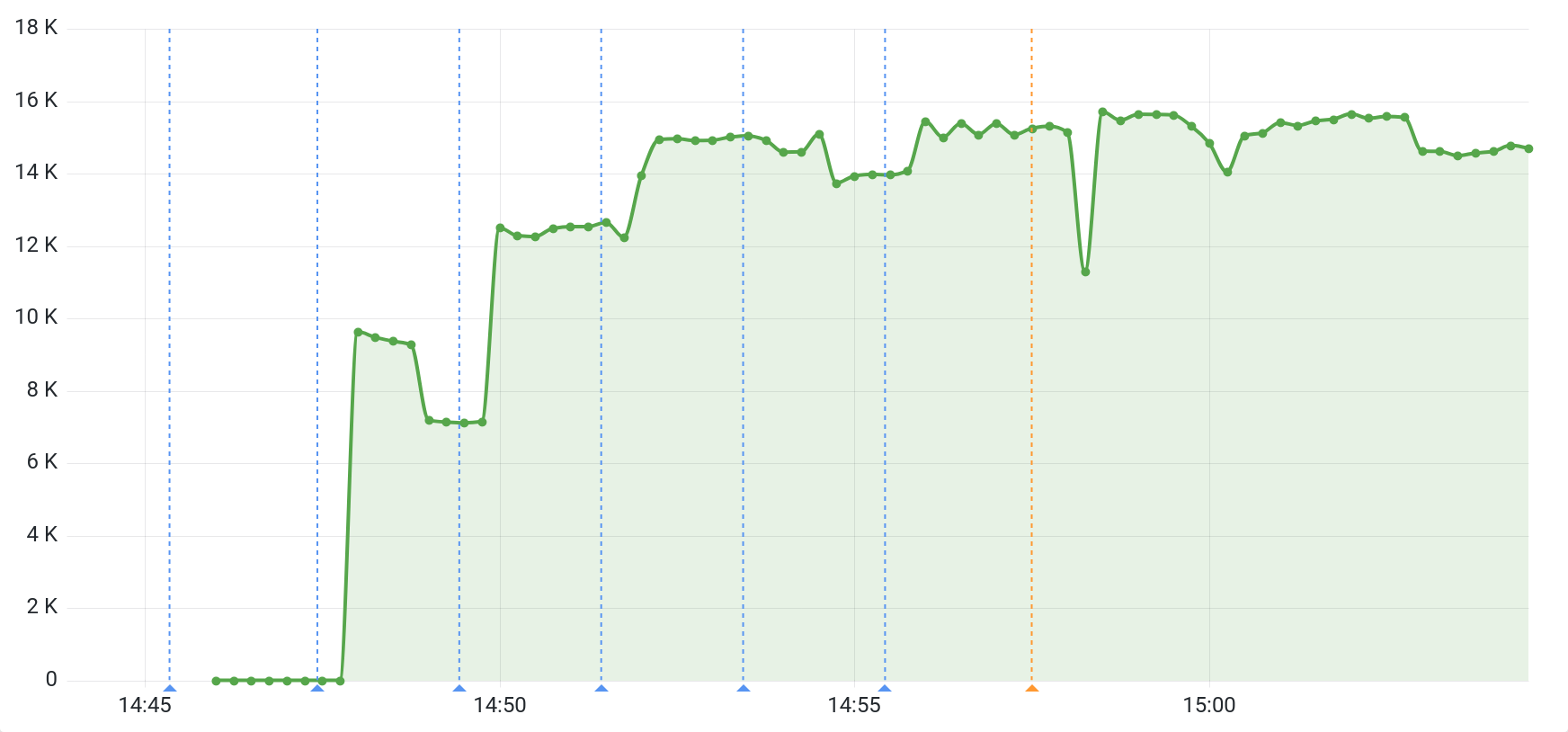}} 
\\ 
\subfloat[Avg. Throughput per client]{\includegraphics[page=1,width=\columnwidth]{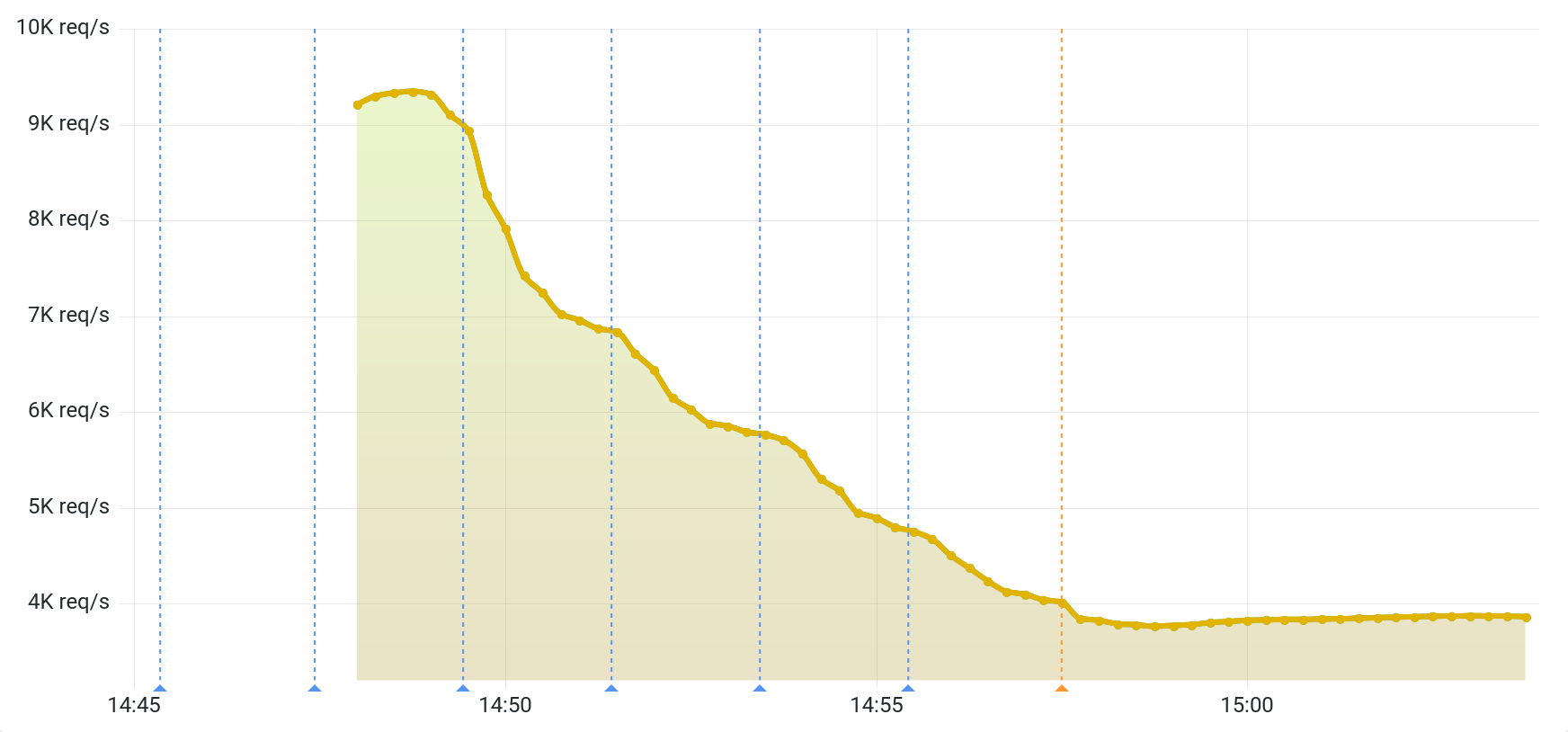}} 
\\ 
\subfloat[Avg. Tail Latency(99.99\%) per client]{\includegraphics[page=1,width=\columnwidth]{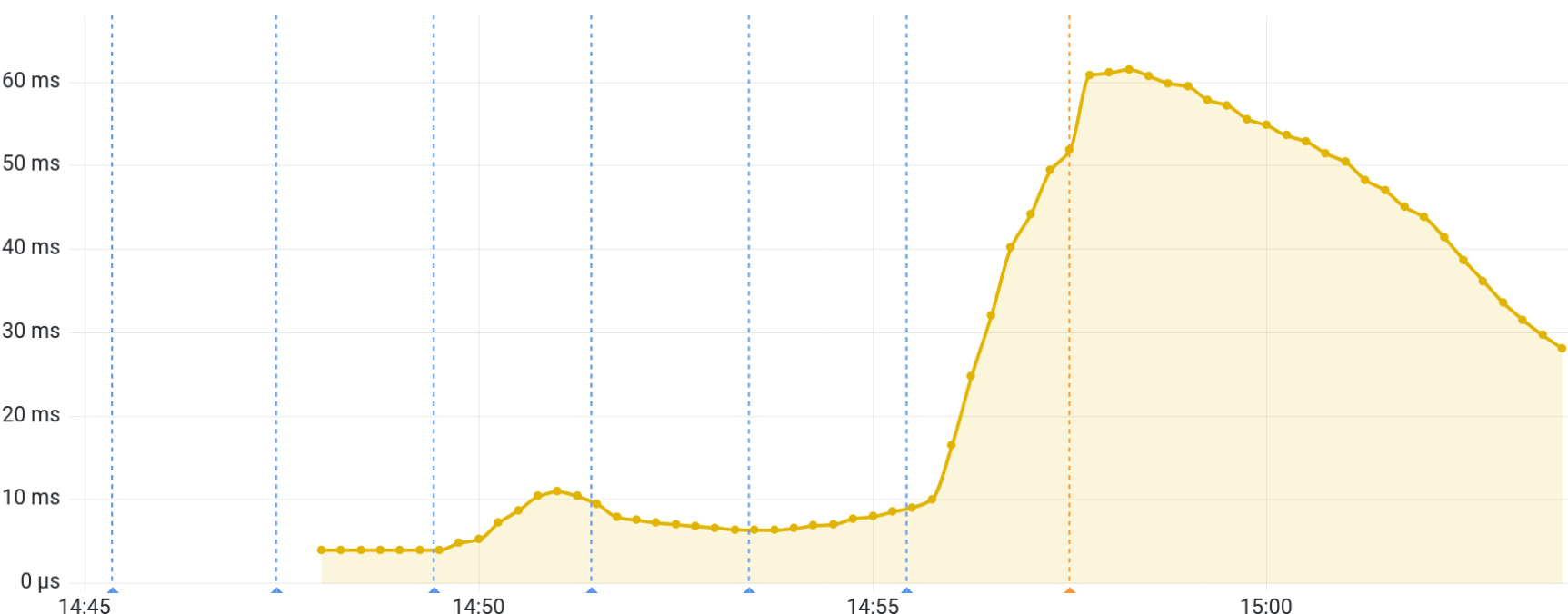}}
\\
\caption{Identify saturation points. The blue lines indicate the time new stressing services were added. The yellow line indicates that \frisbee{} has detected a saturation point and has removed the service causing the excessive load.}
\label{eval:saturation}
\end{figure}

\subsection{Deployment: Saturation test}
In this test, we show how \frisbee{} can find the golden spot at which a Redis server is not too overloaded yet also not too underutilized. This spot, known as saturation point, is especially important for reducing the Total Cost of Ownership (TCO) when deploying a system on cloud platforms that charge based on disk capacity, IOPS, and network traffic. The experiment exercises: (i) performance alerts, (ii) contextualized visualization from different views (master, client), (iii) placement constraints, (iv) scheduling constraints, and (v) resource throttling. The workflow description is 48 lines.

\paragraph{Methodology}
\frisbee{} automates the process mentioned above by gradually increasing request load until some percentage of SLA violations occur. For the request load, we use a logical group that schedules the execution of a new client every two minutes. We claim that we have detected a \emph{saturation point} if, despite the increasing load, the average rate of requests processed by Redis does not increase more than 10\% in a rolling window of three minutes. The respective expression is `WHEN percent\_diff () OF query (avg(redis\_ops\_per\_sec), 1m, now-3m) IS BELOW 10'. By using a window slightly higher than the service creation interval (spacing), we account for the effects of the service just below the saturation point and the service just above it. If there is an alert within that window, the workflow takes two actions. Firstly, it suspends the distributed group so that no more clients are added. Secondly, it invokes a killing action and removes the latest client that saturated the system. If the number of processed requests remains stable after that point, we claim we have found the saturation point. 

\paragraph{Detecting stability points}
Figure~\ref{eval:saturation} shows the number of processed requests on the server (utilization), as well as the average throughput and average tail-latency per client. The first thing is to note that if the rolling window is lower than the creation interval, it is possible to have a low-end stable performance (less than 10\% variation) and mistake it for saturation. Secondly, when the experiment starts, only one client does not fully utilize the server, and therefore benefits from high throughput and low latency. As the number of clients increases, we see a minor increase in server utilization, but a rapid decrease in the average throughput per client -- albeit the total through of all clients remains the same. However, despite the decrease in throughput, the tail latency remains low. After a few services are added, we notice that the rate of change in server utilization and client throughput is relatively low, but there is a rapid increase in the tail latency. At this point, Grafana raises an alert that is subsequently captured by the controller, which removes the last service. After that, the utilization and the throughput remain stable, whereas the tail latency rapidly decreases. Therefore, we are sure we have successfully spotted the saturation point. 

\paragraph{Porting experiment}
A major selling point of Kubernetes is the portable deployment pipeline. Motivated by that, we ran the same experiment on a different hardware. Instead of a server-grade machine (24 cores @2.4GHz, 128 GB DDR3 RAM) we used a Workstation (8 cores @4.2GHz, 32 GB DDR4 RAM). Despite the less cores, we noticed a 3x improvement in the peak throughput (from 16K requests/sec to 42K requests/sec). We attribute this behavior to the fact that Redis is designed around top-level locks, thus benefiting from powerful CPUs but not by numerous CPUs. However, the higher processing rate and the reduced memory compared to the server caused performance instabilities that made it difficult to detect the saturation point automatically. Thus, we had to scale down the experiment by using \frisbee{}'s ability for selective CPU throttling (reducing the traffic generated by each client), (ii) reduced the decision window and the spacing between service creations. Indeed, that allowed us to detect the saturation point successfully, but that required some trial-and-error for tuning the experiment's parameters.

\section{Conclusions}
\label{sec:conclusion}
We presented \frisbee{}, a Kubernetes-native platform for exploring, testing, and benchmarking Cloud-native applications. \frisbee{} removes the hurdles of testing complexity from the users, providing a unified language for modeling various classes of experiments. Using \frisbee{}'s abilities, we extract numerous insights that influence the way we evaluate our applications. Our experiments demonstrate the interplay of:
\begin{enumerate*}[label=(\roman*)]
    \item reusable service templates,
    \item logical grouping,
    \item dependency management,
    \item scheduling constraints,
    \item resource throttling,
    \item failure injection,
    \item informed placement, and 
    \item contextualized visualization of the global application state.
\end{enumerate*}

 The ability for mixed I/O workloads revealed that inserting new keys in a running Redis server can affect collocated clients greater than when updating existing keys. The contextualized visualization revealed that workloads often exhibit artifacts, such as slow operation and operation errors, that aggregated values tend to hide. If not accounted, these artifacts may lead to invalid conclusions and unfair comparisons when reasoning about the performance baseline of a system. The chaos capabilities enabled us to simulate network partitions and study the recovery mechanisms of Redis. Via the global view of the system state that \frisbee{} provides, we studied recovery from the perspective of the master node, the slave node, and the client, and found out that the (YCSB) benchmark does not support failover for Redis. Further, we used \frisbee{}'s ability to perform real-time analysis on performance metrics, which allowed us to discover the saturation point of a Redis server automatically. Finally, in a effort to use the same configuration to find the saturation point on different hardware, \frisbee{} ability to perform resource throttling appeared extremely helpful to scale down the problem to the capacity of the hardware.

\section{Future Work}
\label{sec:future}
We have already integrated \frisbee{} into the continuous testing pipelines of our team to run a repeatable set of tests whenever we make a commit to the code repository. Our future research goals aim in two directions. The first is to improve confidence in the tests by reducing their flakiness. To do that, we want to integrate test analysis tools and automate the reiteration of tests until the margin of error is beyond a given threshold. The second is to extend declarative testing with model-testing aptitudes, to support auto-generated tests that fit into the capacity of a given physical host.


\section{Acknowledgements}
This project has received funding from the European Union's Horizon 2020 research and innovation programme under the Marie Sk\l{}odowska-Curie, grant agreement No. 894204 (Ether, H2020-MSCA-IF-2019).

\section{Availability}
As we believe that platforms like \frisbee{} will play a catalytic role in building robust, recoverable, and performant cloud-native applications, we have released the source code, the templates, and the monitoring packages at: https://github.com/CARV-ICS-FORTH/frisbee.



\bibliographystyle{plain}
\bibliography{ms}

\end{document}